\documentclass{ifacconf}

\usepackage{graphicx}      
\usepackage{natbib}        

\usepackage{amsmath, amsfonts, amssymb, dsfont}
\usepackage{nicefrac}

\newcommand{\vertiii}[1]{{\left\vert\kern-0.25ex\left\vert\kern-0.25ex\left\vert #1 
    \right\vert\kern-0.25ex\right\vert\kern-0.25ex\right\vert}}
\DeclareMathOperator{\supp}{supp}

\usepackage{tikz}
\usetikzlibrary{positioning}

\usepackage{amsmath}
\usepackage{amssymb}
\usepackage[old]{old-arrows}

\makeatletter
\newcommand*{\@tto}[2]{%
   \dimen@\fontdimen8
       \ifx#1\displaystyle\textfont\else
       \ifx#1\textstyle\textfont\else
       \ifx#1\scriptstyle\scriptfont\else
       \scriptscriptfont\fi\fi\fi 3
   \mathrel{%
      \vcenter{%
         \vbox{%
            \baselineskip\z@skip
            \lineskip\z@
            \ialign{##\cr$#1\varrightarrow$\cr
            \noalign{\kern-\dimen@}%
            $#1\varrightarrow$\cr}%
         }%
      }%
   }%
}

\makeatletter
\newcommand{\longtto}{\mathpalette\@longtto\relax}
\newcommand*{\@longtto}[2]{%
   \dimen@\fontdimen8
       \ifx#1\displaystyle\textfont\else
       \ifx#1\textstyle\textfont\else
       \ifx#1\scriptstyle\scriptfont\else
       \scriptscriptfont\fi\fi\fi 3
   \mathrel{%
      \vcenter{%
         \vbox{%
            \baselineskip\z@skip
            \lineskip\z@
            \ialign{##\cr$#1\varlongrightarrow$\cr
            \noalign{\kern-\dimen@}%
            $#1\varlongrightarrow$\cr}%
         }%
      }%
   }%
}

\makeatletter
\newcommand{\mapstto}{\mathpalette\@mapstto\relax}
\newcommand*{\@mapstto}[2]{%
   \dimen@\fontdimen8
       \ifx#1\displaystyle\textfont\else
       \ifx#1\textstyle\textfont\else
       \ifx#1\scriptstyle\scriptfont\else
       \scriptscriptfont\fi\fi\fi 3
   \mathrel{%
      \vcenter{%
         \vbox{%
            \baselineskip\z@skip
            \lineskip\z@
            \ialign{##\cr$#1\mapstochar\varrightarrow$\cr
            \noalign{\kern-\dimen@}%
            $#1\mapstochar\varrightarrow$\cr}%
         }%
      }%
   }%
}

\makeatletter
\newcommand{\longmapstto}{\mathpalette\@longmapstto\relax}
\newcommand*{\@longmapstto}[2]{%
   \dimen@\fontdimen8
       \ifx#1\displaystyle\textfont\else
       \ifx#1\textstyle\textfont\else
       \ifx#1\scriptstyle\scriptfont\else
       \scriptscriptfont\fi\fi\fi 3
   \mathrel{%
      \vcenter{%
         \vbox{%
            \baselineskip\z@skip
            \lineskip\z@
            \ialign{##\cr$#1\mapstochar\varlongrightarrow$\cr
            \noalign{\kern-\dimen@}%
            $#1\mapstochar\varlongrightarrow$\cr}%
         }%
      }%
   }%
}

\begin{document}
\begin{frontmatter}

\title{On Network Congestion Reduction Using Public Signals Under Boundedly Rational User Equilibria (Full Version)\thanksref{footnoteinfo}} 

\thanks[footnoteinfo]{This work was supported by the \emph{ARO} MURI grant W911NF-20-0252 (76582 NSMUR).}

\author[First]{Olivier Massicot} 
\author[Second]{C\'edric Langbort} 

\address[First]{Coordinated Science Lab, Urbana, IL 61801 (e-mail: om3@illinois.edu).}
\address[Second]{Coordinated Science Lab, Urbana, IL 61801 (e-mail: langbort@illinois.edu).}

\begin{abstract}
Boundedly Rational User Equilibria (BRUE) capture situations where all agents on a transportation network are electing the fastest option up to some time indifference, and serve as a relaxation of User Equilibria (UE), where each agent exactly minimizes their travel time. We study how the social cost under BRUE departs from that of UE in the context of static demand and stochastic costs, along with the implications of BRUE on the optimal signaling scheme of a benevolent central planner. We show that the average excess time is sublinear in the maximum time indifference of the agents, though such aggregate may hide disparity between populations and the sublinearity constant depends on the topology of the network. Regarding the design of public signals, even though in the limit where agents are totally indifferent, it is optimal to not reveal any information, there is in general no trend in how much information is optimally disclosed to agents. What is more, an increase in information disclosed may either harm or benefit agents as a whole.
\end{abstract}


\end{frontmatter}

\section{Introduction}

Selfish routing, first erected as a principle by \citet{wardrop1952road}, is a canonical description of how a crowd of self-minded drivers minimizing their own travel time decide on their paths across a road network prone to congestion. The resulting equilibria, termed User Equilibria (UE), stem from independent non-cooperative actions, yet, as \citet{beckmann1956studies} revealed, can be captured as minima of a convex function: the potential of the game \citep{monderer1996potential,sandholm2001potential}. This fact notoriously implies their existence and (essential) uniqueness. The social cost induced by a UE, i.e., the sum of travel times, is higher than the social optimum, where the itinerary of all vehicles could be chosen by a benevolent central planner. Several attempts such as \citet{cole2003pricing} have been made to alleviate congestion by changing the incentives of drivers through pricing. In this article, however, we focus on the softer option of shaping utilities through public information provision. 

Such information provision can be helpful, in particular in a network subject to random events such as lane closures, accidents, traffic jams, and so on, since a crowd of uninformed drivers might make inefficient decisions. In fact, without any form of signaling, the social cost under such situation may be arbitrarily high compared to that where drivers know the state of the network perfectly \citep{massicot2021competitive}. On the other hand, indiscriminate revelation of all available information to all drivers may also trigger social inefficiencies. For example, if drivers are fully informed they may congest clear roads, while it may be more socially desirable that a fraction of them take a longer route.

In order to determine efficient information revelation policies (which are typically more complex than full revelation or obfuscation) that a central planner can implement to alleviate congestion in a stochastic road network with use of public signals, a number of previous works such as \citet{das2017reducing,zhu2018stability,zhu2019routing,zhu2022information,massicot2019public,massicot2021competitive,castiglioni2021signaling,ferguson2022avoiding} have relied of the notion and tools of Bayesian Persuasion first introduced by \citet{Kamenica}. However, these studies all fall short of considering the empirical fact (confirmed for instance by \citet{avineri2004violations,morikawa2005driver}) that drivers often fail to exactly minimize their travel time. Although a plethora of models is available in the transportation literature, each most adapted to a specific situation or effect (see \citet{di2016boundedly} for a review), the specific indifference bands model of \citet{di2013boundedly}, reminiscent of almost-best-responding agents found in economics \citep{dixon2017surfing}, where drivers choose a path minimizing their travel time up to some preset $\epsilon\geq0$ (termed $\epsilon$-BRUE for Boundedly Rational User Equilibria in this article) remains the simplest relaxation of UE.

In parallel to the literature relaxing the time-minimizing assumption in modeling the choices of a crowd, a recent branch of the Bayesian persuasion literature (such as \citet{tang2021bayesian,ball2021experimental,de2022non}) strives to relax the traditional assumptions that the receiver is a Bayesian expected utility maximizer. Recently, a most simple model has emerged in view of persuading learning agents \citep{lin2024persuading}. Their central assumption has the receiver taking any of the actions that offer a payoff at most $\epsilon\geq0$ less than the optimal one, and as such, is an exact equivalent of that of \citet{di2013boundedly} used in modeling route choice.

In this article, we explore how public signals may be used to alleviate the congestion in non-atomic stochastic congestion games with static demand, where drivers are merely assumed to choose a path at most $\epsilon$ suboptimal. Our article is thus, to the best of our knowledge, the first attempt to conjure robust behavioral models for information design in large games. 

Incidentally, we reveal the novel fact, pertaining to route choice in all generality (and not narrowly to information design), that the difference between the social cost of an $\epsilon$-BRUE and that of a UE is sublinear in $\epsilon$. In other words, if drivers are content with an $\epsilon$ suboptimal choice, in average they may only spend an additional time multiple of $\epsilon$ compared to the UE situation. The multiplicative factor for small $\epsilon$ turns out to depend on the topology of the network. This stands in stark contrast with the price of anarchy results of \citet{roughgarden2002price} by exhibiting topology effects on equilibria inefficiency. We show this result extends to equilibria under public signals, thereby providing a suboptimality guarantee of optimal information disclosing mechanisms under $\epsilon$-BRUE for small $\epsilon$. For larger $\epsilon$, however, the central planner may wish to tackle the problem exactly. We find that in the limit of large $\epsilon$, it is optimal to not reveal any information to the drivers. On the other hand, there are cases in which the amount of information---measured using the Blackwell order \citep{blackwell1953equivalent}---optimally disclosed assuming an $\epsilon$-BRUE is played out increases with $\epsilon$. Finally, we exhibit an instance in which no monotonicity is observed.

The article is structured as follows. In Section \ref{sec:formal}, we lay the formal model: the static network model, the notion of UE, and the notion of $\epsilon$-BRUE as introduced by \citet{di2013boundedly}. Section \ref{sec:welfare} investigates the effects of bounded rationality on social cost. We show the price of anarchy is ill-adapted to quantify the inefficiency caused by bounded rationality and propose our own metric, the average excess time, for which we present an upper bound. Section \ref{sec:public} aims at designing a signaling scheme to alleviate congestion. We extend the previous bound to stochastic networks, allowing the planner to use UE flows for design, granted drivers play an $\epsilon$-BRUE with $\epsilon$ small enough. For larger $\epsilon$, the planner may wish to solve the exact problem; we study how the optimal robust policy varies with $\epsilon$. 

We have relegated all formal proofs to the appendix to remain concise---a brief intuition is nonetheless provided in-text whenever relevant.

\section{Formal model}
\label{sec:formal}

\subsection{Road networks and User Equilibria}

We consider a directed graph $\mathcal G = (\mathcal V, \mathcal E)$ whose vertices represent road intersections and edges road sections, a nonempty set of nonvacuous trips $\mathcal T \subset \mathcal V^2$ and a demand $d\in(\mathbb R_+^*)^{\mathcal T}$ (i.e., a positive vector indexed by trips). For each trip $t=(a,b)\in\mathcal T$, we denote by $\mathcal P_t$ the set of simple paths that join $a$ to $b$, assumed nonempty, and $\mathcal P = \bigsqcup_{t\in\mathcal T} \mathcal P_t$. A flow on paths $f\in(\mathbb R_+)^\mathcal P$ satisfies demand $d$ if for all trip $t\in\mathcal T$,
\[
	\sum_{P\in\mathcal P_t} f_P = d_t.
\]
We denote by $\mathcal F_d$ the set of all such flows. A flow on paths $f\in\mathcal F_d$ induces a flow on edges, still denoted $f$ by overloading the notation, defined by,
\[
	f_e = \sum_{P\ni e} f_P,
\]
where $P\ni e$ tacitly refers to $P\in\mathcal P$ such that $e\in P$. We endow each edge $e\in\mathcal E$ with a latency function $c_e \colon \mathbb R_+ \to \mathbb R_+$, assumed continuous and nondecreasing. The travel time of edge $e\in\mathcal E$ is then $c_e(f_e)$, and the travel time of path $P\in\mathcal P$ is, overloading the notation once more,
\[
	c_P(f) = \sum_{e\in P} c_e(f_e).
\]
We will call the triple $(\mathcal G,c,d)$, a \emph{network}. The usual metric that defines the quality of a flow $f\in\mathcal F_d$ is the social cost, defined as the `sum of all travel times', i.e., formally by 
\[
	\Psi(f) = \sum_{P\in\mathcal P} f_P c_P(f) = \sum_{e\in\mathcal E} f_e c_e(f_e).
\]

The common assumption in studying such model is that selfish drivers trying to minimize their own travel time ultimately route according to a User Equilibrium (later referred to as UE). 
\begin{defn}
	A flow $f\in\mathcal F_d$ is said to be a \emph{User Equilibrium (UE)} of a given network $(\mathcal G,c,d)$, if for all trip $t\in\mathcal T$ and paths $P,Q\in\mathcal P_t$ realizing this trip, $P$ is chosen by some drivers (i.e., if $f_P>0$) only if $P$ is at most as long as $Q$, namely, $c_P(f) \leq c_Q(f)$. 
\end{defn}

\citet{beckmann1956studies} have shown the remarkable fact that such equilibria can be captured exactly as minima of the convex potential
\[
	\Phi(f) = \sum_{e\in\mathcal E} \int_0^{f_e} c_e(x)\,\mathrm dx.
\]
This establishes not only the existence of UE but their essential uniqueness as well, in the sense that if $f,g\in\mathcal F_d$ are UE, then $c_e(f_e) = c_e(g_e)$ for all $e\in\mathcal E$. In particular, the social cost of all UE is identical, we will denote it $\Psi_0$.

This notion of equilibrium is satisfying in many ways: it makes a unique prediction as to what flow (or at least travel time) should be observed on each edge, computing such flows resorts to a convex program, and their worst-case inefficiency relative to the optimal flow (the price of anarchy) is well-characterized. On the other hand, this definition requires agents to act perfectly, whereas it has been found empirically that drivers rather choose a `satisfiable' path. We explore this possibility in the next subsection.

\subsection{Boundedly Rational User Equilibria}

If instead of choosing shortest paths, drivers are satisfied with any path within $\epsilon\geq0$ of the optimal time, they are termed $\epsilon$-utility maximizers (where their utility is the opposite of their travel time) and their aggregate behavior is termed an $\epsilon$-BRUE, standing for Boundedly Rational User Equilibrium.

\begin{defn}
	A flow $f\in\mathcal F_d$ is said to be an \emph{$\epsilon$-BRUE} of a given network $(\mathcal G,c,d)$, if for all trip $t\in\mathcal T$ and paths $P,Q\in\mathcal P_t$ realizing this trip, $P$ is chosen by some drivers (i.e., if $f_P>0$) only if $c_P(f) \leq c_Q(f) + \epsilon$. 
\end{defn}

It is convenient to introduce the threshold function $\varepsilon \colon \mathcal F_d \to \mathbb R_+$ where
\[
	\varepsilon(g) = \max_{\substack{t\in\mathcal T\\P,Q\in\mathcal P_t\\\text{s.t. } g_P>0}} ~c_P(g) - c_Q(g), 
\]
defined precisely so that $g$ is an $\epsilon$-BRUE if and only if $\epsilon\geq\varepsilon(g)$. This function also serves in defining the worst-case social cost under an $\epsilon$-BRUE flow:
\begin{equation}
\label{eq:psiepsilon}
	\Psi_\epsilon \triangleq \sup_{\substack{g\in\mathcal F_d\\\varepsilon(g)\leq\epsilon}} ~\Psi(g).
\end{equation}

\begin{prop}
\label{lem:varepsilon}
	Given a network $(\mathcal G,c,d)$, $g\in\mathcal F_d$ is an $\epsilon$-BRUE if and only if $\epsilon\geq\varepsilon(g)$. Moreover, the function $\varepsilon$ is lower semicontinuous and $\epsilon\mapsto\Psi_\epsilon$ is upper semicontinuous, nondecreasing and ultimately constant.
\end{prop}

Lastly, we note the following variational inequality, well-known when $\epsilon=0$. 

\begin{prop}
\label{lem:variational}
	Given a network $(\mathcal G,c,d)$, a flow $f\in\mathcal F_d$ is an $\epsilon$-BRUE if and only if for all $g\in\mathcal F_d$,
	\[
		\sum_{P\in\mathcal P} (f_P-g_P) c_P(f) \leq \frac12\epsilon\sum_{P\in\mathcal P} |f_P-g_P| = \frac12\epsilon \|f-g\|_1.
	\]
\end{prop}

This inequality notably implies that, just like UE are minima of $\Phi$, $\epsilon$-BRUE are almost minimizing $\Phi$. More precisely, we can show the following proposition.
\begin{prop}
\label{lem:variational2}
	Given a network $(\mathcal G,c,d)$ and flow $f\in\mathcal F_d$, denoting $\Phi_0 = \min_{\mathcal F_d} ~\Phi$, 
	\[
		\Phi(f) \leq \Phi_0 + \|d\|_1 \varepsilon(f).
	\]
\end{prop}
Here $\|d\|_1$ refers to the $1$-norm of $d$, i.e., the total demand. It is important to note, however, that almost-minimizers of $\Phi$ are not necessarily $\epsilon$-BRUE as the following example illustrates.

\begin{exmp}
	Consider the unit-demand two-road network with constant costs $1$ and $2$. The flow $(1,0)$ is the unique $\epsilon$-BRUE when $\epsilon<1$, whereas the set of $\epsilon$-minimizers of $\Phi$ is all $(1-x,x)$ with $x\in[0,\epsilon]$.
\end{exmp}

\section{Welfare effect of bounded rationality}
\label{sec:welfare}

\subsection{Price of anarchy}

\citet{roughgarden2005selfish} has found that, although road congestion is an example of the tragedy of the commons where the selfish action of each agent deteriorates the overall conditions, the price of anarchy---the worst-case ratio, for all network, of social cost under UE over the minimal one---is actually bounded. One could envisage extending this result to $\epsilon$-BRUE, however this notion makes little sense in this context as the following example shows. 
\begin{exmp}
	Consider a unit-demand two-road network with constant costs $\epsilon\in(0,1)$ and $\epsilon^2$ respectively for the first and second road. The flow $(1,0)$ is an $\epsilon$-BRUE whereas the flow $(0,1)$ minimizes the social cost, hence the price of anarchy of this instance is $\epsilon^{-1} = \nicefrac\epsilon{\epsilon^2}$, which is unbounded as $\epsilon$ vanishes.
\end{exmp}

Nonetheless, in this specific example, a meaningful \emph{additive} comparison of the social cost under $\epsilon$-BRUE and UE can be drawn: $\Psi_\epsilon \leq \Psi_0 + \epsilon$. In other words, for a flow $g\in\mathcal F_d$, the \emph{average excess time}
\[
	\psi(g) \triangleq \frac{\Psi(g)-\Psi_0}{\|d\|_1}
\]
is upper-bounded by $\varepsilon(g)$. We investigate further whether such statement can be generalized.

\subsection{Average excess time due to bounded rationality}

We first note the previous remark holds for all networks of constant costs: the travel times do not depend on the flows hence the path taken by drivers under an $\epsilon$-BRUE are at most $\epsilon$ longer than the shortest paths. More precisely, for such networks, $\psi \leq \varepsilon$, namely for all flow $g\in\mathcal F_d$,
	\begin{equation}
	\label{eq:constantcosts}
		\Psi(g) \leq \Psi_0 + \|d\|_1 \varepsilon(g).
	\end{equation}
Upon closer inspection, this fact can be shown to also hold for all parallel-road networks.

\tikzset{
    vertex/.style={circle,draw,minimum size=10pt,outer sep=0pt},
    block/.style={rectangle,draw,minimum size=10pt,outer sep=0pt,rounded corners=2pt}
}

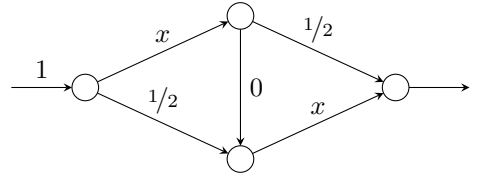
\begin{figure}
\centering
\begin{tikzpicture}[	>=stealth,
								node distance=0.7cm and 1.8cm]
	\node[vertex] (A) {};
	\node[left=0.8cm of A] (s) {};
	\node[vertex, above right=of A] (B) {};
	\node[vertex, below right=of A] (C) {};
	\node[vertex, below right=of B] (D) {};
	\node[right=0.8cm of D] (t) {};
	
	\draw[->] (s) -- (A) node[midway, above] {$1$};
	\draw[->] (D) -- (t);
	\draw[->] (A) -- (B) node[midway, above] {$x$};
	\draw[->] (B) -- (C) node[midway, right] {$0$};
	\draw[->] (C) -- (D) node[midway, above] {$x$};
	\draw[->] (A) -- (C) node[midway, above] {$\nicefrac12$};
	\draw[->] (B) -- (D) node[midway, above] {$\nicefrac12$};
\end{tikzpicture}
\caption{A Wheatstone network, with unit demand crossing from the left to the right.}
\label{fig:wheatstone}
\end{figure}

However a richer example reveals that \eqref{eq:constantcosts} cannot stand as-is for an arbitrary network. 
Indeed, consider the Wheatstone network pictured in Figure \ref{fig:wheatstone}, with unit demand crossing from the left node to the right one. The dummy variable $x$ refers to the flow traversing each edge and the edge labels refer to the cost function. We call $U,M,D$ the up, middle and down paths. There is a unique UE, $f=(\nicefrac12,0,\nicefrac12)$, generating cost $1$ on each of the three paths. Consider the flow $g=(0,1,0)$, the costs under this flow are
\[
	c_U(g) = \frac32, ~c_M(g) = 2, ~c_D(g) = \frac32.
\]
As a result, $\varepsilon(g) = \nicefrac12$ and
\[
	\psi(g) = \Psi(g) - \Psi(f) = 1 = 2\varepsilon(g) > \varepsilon(g),
\]
which invalidates \eqref{eq:constantcosts}. Nonetheless, in this specific example, one can show by separating each case based on $h$'s support that for all $h\in\mathcal F_d$, $\psi(h) \leq 2\varepsilon(h)$, and so, even though the average excess time exceeds $\varepsilon(h)$, it remains a mere multiple of it. This suggest a slightly different right-hand side term in \eqref{eq:constantcosts}, which the following theorem establishes in full generality.

\begin{thm}
\label{thm:continuity}
	Given a directed graph $\mathcal G$, there exists a constant $L_{\mathcal G}\geq0$ such that for all network $(\mathcal G, c, d)$,
	\[
		\psi(g) \leq L_{\mathcal G} \varepsilon(g) + o(\varepsilon(g)).
	\]
	In turn, fixing a network $(\mathcal G,c,d)$, there exists a constant $M_{(\mathcal G,c,d)}\geq0$ such that for all $g\in\mathcal F_d$,
	\[
		\psi(g) \leq M_{(\mathcal G,c,d)} \varepsilon(g).
	\]
\end{thm}

We let $\mathcal L_{\mathcal G}$ and $\mathcal M_{(\mathcal G,c,d)}$ respectively denote the smallest $L_{\mathcal G}$ and $M_{(\mathcal G,c,d)}$ such that the statements of Theorem \ref{thm:continuity} holds. The inefficiency factor $\mathcal L_{\mathcal G}$ possibly depends on the topology of the network though not on the costs and demand, whereas the factor $\mathcal M_{(\mathcal G,c,d)}$ could a priori depend on the demand and costs as well. In the next subsection, we investigate whether $\mathcal L_{\mathcal G}$ is unbounded over all graphs $\mathcal G$, i.e., whether a universal constant $\mathcal L$ exists.

\subsection{Unboundedness of the inefficiency factor}

\begin{figure}
\centering
\begin{tikzpicture}[	>=stealth,
								node distance=1.5cm and 1.5cm]
	\node[block] (W1) {$W_1$};
	\node[left=0.8cm of W1] (s) {};
	\node[block, right=of W1] (W2) {$W_2$};
	\node[right=0.5cm of W2] (d) {$\ldots$};
	\node[block, right=of W2] (WN) {$W_N$};
	\node[right=0.8cm of WN] (t) {};
	
	\draw[->] (s) -- (W1);
	\draw[->] (WN) -- (t);
	\draw[->] (W1) -- (W2);
	\draw[->] (W2) -- (d);
	\draw[->] (d) -- (WN);
\end{tikzpicture}
\caption{A chain of Wheatstone networks identical to those of Figure \ref{fig:wheatstone}.}
\label{fig:wheatstones}
\end{figure}
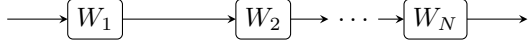

Chaining $N$ identical copies of the Wheatstone network of Figure \ref{fig:wheatstone}, as depicted in Figure \ref{fig:wheatstones}, we arrive at a network of inefficiency factor $\mathcal L_{\mathcal G} \geq N$; this result is stated in the next proposition.

\begin{prop}
\label{prop:LG}
	For all $N$ positive integer, there exists a directed graph $\mathcal G$ with $3N+1$ vertices and $5N$ edges such that $\mathcal L_{\mathcal G} \geq N$.
\end{prop}

The inefficiency factor $\mathcal L_{\mathcal G}$ thus truly depends on the topology of the network. This result is in striking contrast with those concerning the price of anarchy. Indeed, the price of anarchy has been shown by \citet{roughgarden2005selfish} to only depend on the costs, not on the topology, whereas the inefficiency factor depends on the topology but not on the costs. 

The instance used in the proof of Proposition \ref{prop:LG}, presented in the following example, also illustrates how averaged inefficiency inequalities such as $\psi \leq M \varepsilon$ or $\psi \leq L \varepsilon + o(\varepsilon)$ may hide important disparities among drivers.

\begin{exmp}
Take the network pictured in Figure \ref{fig:wheatstones} with demand $d' \in (0,1)$ on each subnetwork and $1-d'$ to cross the entire network. The flow $f$ where all drivers choose only $U$ and $D$ paths with equal proportion is a UE since the cost of the $U,M,D$ paths on each subnetwork is $1,1,1$. On the other hand, the flow $g$ identical to $f$ save for drivers on each subnetwork instead taking path $M$ is an $\epsilon$-BRUE where $\epsilon=\nicefrac{d'}2$, as the cost of the subnetwork paths are now $1+\epsilon,1+2\epsilon,1+\epsilon$. As a result of their $\epsilon$-bounded rationality, the drivers on the subnetworks experience an excess time of $2\epsilon$ at $g$ compared to the situation at UE, whereas the agents crossing the entire network suffer an excess time of $N\epsilon$, in spite of choosing the shortest path available to them.
\end{exmp}

\section{Information provision for bounded rationality}
\label{sec:public}

\subsection{Stochastic network}

In order to model random events which may affect network conditions, we consider the following enrichment of deterministic networks. A \emph{stochastic network} is a quadruple $(\mathcal G, c, d, \nu)$, where the costs $c$ depend on a random variable denoted $\omega$ and distributed according to the probability measure $\nu$ (the probability space $\Omega$ is assumed finite, with $\sigma$-algebra $2^\Omega$, and kept implicit in the notation). 

When information about the network is revealed to all agents at once via a public signal, they may hold a common posterior belief $\mu$, and the network thus appears to them (if they are $\epsilon$-utility maximizers with $\epsilon\geq0$) as a deterministic network $(\mathcal G, c^\mu, d)$, with expected edge costs $c_e^\mu(f_e) = \mathbb E_\mu[c_e(f_e,\omega)]$. The notions of UE and BRUE under common belief $\mu$ are thus defined much like their deterministic counterpart, albeit with expected travel times. When $\mu$ is a belief, we will then use the exponent $\mu$ to denote the attributes of the network $(\mathcal G, c^\mu, d)$ such as social costs $\Psi^\mu_\epsilon$ and threshold function $\varepsilon^\mu$.

\subsection{Information design}

Since the information drivers have access to shapes their utility, it is possible in principle for a central planner who observes $\omega$ to nudge the crowd of agents by partially revealing it. The messaging policy is typically a random variable correlated with $\omega$. This opportunity was first realized by \citet{das2017reducing} who make use of Bayesian persuasion \citep{Kamenica}, which defines and solves this very problem in the case of a single agent. It is possible to apply the same tools to the problem at hand, having first defined the objective of the central planner. In this article, we consider that the central planner wishes to minimize the expectation of $\Psi_\epsilon^\mu$ (defined as \eqref{eq:psiepsilon}), the worst-case social cost assuming drivers will route according to an $\epsilon$-BRUE. 

More specifically, the problem of the planner can be broken down into two steps: 1) finding a distribution $\tau$ of posterior beliefs induced by a messaging policy that minimizes the expected worst-case social cost, $\mathbb E_\tau[\Psi^\mu_\epsilon]$, and 2) retrieving a signaling policy that induces such an optimal distribution $\tau$. \citet{Kamenica} have established that 1) a distribution of belief $\tau$ can be induced by a policy if and only if it is Bayes-plausible, i.e., if and only if $\mathbb E_\tau[\mu] = \nu$, 2) there exists an optimal $\tau$ with support of cardinal lesser than or equal to $|\Omega|$, the number of events. In other words, the first search boils down to the following program where $\tau$ is any distribution of beliefs with support of cardinal $|\Omega|$ at most,
\[
	\begin{aligned}
		\min_{\tau} &\quad\mathbb E_\tau[\Psi^\mu_\epsilon] \\
		\text{s.t.} &\quad\mathbb E_\tau[\mu] = \nu,
	\end{aligned}
\]
whereas the second step is a mere inverse problem involving Bayes' formula. Solving this program turns out to be challenging as evaluating $\Psi^\mu_\epsilon$ is computationally hard \citep{di2013boundedly}. Hence, in the next subsections, we discuss possible practical alternatives as well as qualitative results.

\subsection{Designing information using UE as a proxy}

In contrast with $\Psi^\mu_\epsilon$, $\Psi^\mu_0$ is relatively amenable to computation since it is merely $\Psi^\mu(f)$ where $f$ is a solution to the convex program $\min_{\mathcal F_d} ~\Phi^\mu$. For this reason, a central planner might be satisfied with the guarantees granted by the generalization of Theorem \ref{thm:continuity} stated below, and choose to design a signal assuming drivers are exact time-minimizers. 

\begin{cor}[of Theorem \ref{thm:continuity}]
\label{cor:continuity}
	Given a stochastic network $(\mathcal G,c,d,\nu)$ and a finite distribution $\tau$ of posterior beliefs $\mu$, 
	\[
		\mathbb E_\tau[\psi^\mu(g^\mu)] \leq \mathcal L_{\mathcal G} \mathbb E_\tau[\varepsilon^\mu(g^\mu)] + o(\mathbb E_\tau[\varepsilon^\mu(g^\mu)]),
	\]
	where for all $\mu\in\supp\tau$, $g^\mu\in\mathcal F_d$.
\end{cor}

In the case where $\epsilon$ is not insignificant, the central planner may however wish to design a policy so as to exactly minimize the expectation of $\Psi_\epsilon^\mu$. As this remains challenging, we now explore several instances where we can nonetheless exhibit the optimal policy.

\subsection{Designing information for large $\epsilon$}

In the limit case of large $\epsilon$, the constraint $\varepsilon(g) \leq \epsilon$ is not binding anymore in \eqref{eq:psiepsilon}, $\mu\mapsto\Psi^\mu_\epsilon$ becomes convex, and thus not revealing any information becomes optimal. This case is the subject of the following proposition.

\begin{prop}
\label{lem:epsilonlarge}
	Given a stochastic network $(\mathcal G, c, d, \nu)$, for all $\epsilon\geq\epsilon_\infty$ where
\[
	\epsilon_{\infty}=\max_{\substack{P\in\mathcal P\\f\in\mathcal F_d\\\omega\in\Omega}}~c^\omega_P(f),
\]
	it is optimal to not reveal any information to minimize the expectation of $\Psi^\mu_\epsilon$.
\end{prop}

One may thus wonder whether the decrease in information revelation by an optimal policy, as measured by the Blackwell order \citep{blackwell1953equivalent}, is monotonous in $\epsilon$. As it turns out, the answer to this question is negative, there are simple instances where the amount of disclosed information instead increases. 

\subsection{Information revealed may increase with $\epsilon$}

\begin{figure}
\centering
\begin{tikzpicture}[	>=stealth,
								node distance=1.5cm and 2.5cm]
	\node[vertex] (A) {};
	\node[left=0.8cm of A] (s) {};
	\node[vertex, right=of A] (B) {};
	\node[right=0.8cm of B] (t) {};
	
	\draw[->] (s) -- (A) node[midway, above] {$1$};
	\draw[->] (B) -- (t);
	\draw[->] (A) to [bend right] node[midway, above] {$\nicefrac x2+\nicefrac32$} (B);
	\draw[->] (A) to [bend left] node[midway, above] {$\nicefrac x5+\nicefrac85 ; 2$} (B);
\end{tikzpicture}
\caption{A stochastic network for which information revealed increases with $\epsilon$, yet $\epsilon_0$-minimizer agents are worse off if the central planner believes $\epsilon\geq\epsilon_0$.}
\label{fig:worseoff}
\end{figure}
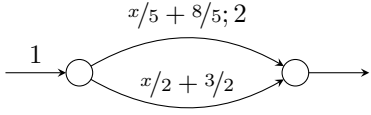

Consider $\mathcal N_1$, the two-road network of Figure \ref{fig:worseoff} with $\omega$ binary deciding whether the cost of the upper road is $\nicefrac x5+\nicefrac85$ or $2$ (when $\omega=0$ or $1$), and $\nu$ uniform. To craft an optimal signaling policy, we first compute $\Psi_\epsilon^\mu$ for all beliefs $\mu\in[0,1]$ (assimilated to the probability assigned to the event $\omega=1$), then use the concavification procedure pioneered by \citet{Kamenica}. One finds, for $\epsilon\in[0,\nicefrac1{10}]$, 
\[
	\Psi_\epsilon^\mu
	= \min\!\left(2,\frac{12 + \epsilon + 10 \epsilon^2 - 2 \mu + 4 \epsilon \mu}{7 - 2 \mu}\right).
\]
The following lemma makes use of this expression to derive the optimal signaling policy.

\begin{lem}
\label{lem:increase}
	For the stochastic network $\mathcal N_1$, to minimize the expectation of $\Psi^\mu_\epsilon$ it is uniquely optimal to:
	\begin{enumerate}
		\item not reveal any information when $\epsilon\in[0,\nicefrac1{35}]$,
		\item use the signaling policy inducing the posterior beliefs $\delta_1$ and $p(\epsilon)\delta_1 + (1-p(\epsilon))\delta_0$ where
		\[
			p(\epsilon) = {1 - \nicefrac52 \epsilon - \nicefrac52 \sqrt{\epsilon + \epsilon^2}},
		\]
		when $\epsilon\in(\nicefrac1{35},\nicefrac4{45})$,
		\item reveal $\omega$ itself when $\epsilon\in[\nicefrac4{45},\nicefrac1{10}]$.
	\end{enumerate}
	Moreover, 
	$p \colon [\nicefrac1{35},\nicefrac4{45}] \to [0,\nicefrac12]$ is a decreasing bijection, therefore the convex hull of the posterior beliefs increases for the inclusion order. In turn, the information revealed increases in the sense of Blackwell with $\epsilon\in[0,\nicefrac1{10}]$.
\end{lem}

A perhaps surprising fact is that even though more information is shared, the crowd may be worse off. Consider the drivers actually route according to the worst-case $\epsilon_0$-BRUE in all cases where $\epsilon_0\in[0,\nicefrac1{10})$ (this include the UE), so that the true conditional social cost under posterior belief $\mu$ is $\Psi^\mu_{\epsilon_0}$. When $\epsilon\in[\epsilon_0,\nicefrac1{10}]$ increases, the information revealed under the assumption of $\epsilon$-BRUE increases, yet the expectation of the actual social cost increases as well. Such counterintuitive phenomenon has also been noted for UE by \citet{acemoglu2018informational}.

\subsection{Information revealed may be non-monotonic in $\epsilon$}

\begin{figure}
\centering
\begin{tikzpicture}[	>=stealth,
								node distance=1.5cm and 2.5cm]
	\node[vertex] (A) {};
	\node[left=0.8cm of A] (s) {};
	\node[vertex, right=of A] (B) {};
	\node[right=0.8cm of B] (t) {};
	
	\draw[->] (s) -- (A) node[midway, above] {$1$};
	\draw[->] (B) -- (t);
	\draw[->] (A) to [bend right] node[midway, above] {$x+\nicefrac12$} (B);
	\draw[->] (A) to [bend left] node[midway, above] {$x+\omega$} (B);
\end{tikzpicture}
\caption{A stochastic network for which no monotonic (in $\epsilon$) sequence of public signals exist, $\omega\in\{0,1\}$.}
\label{fig:nonmonotonic}
\end{figure}
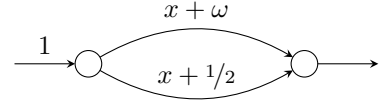

As a final example, consider $\mathcal N_2$ the two-road network of Figure \ref{fig:nonmonotonic} where the top road may be subject to an unknown delay modeled by $\omega\in\{0,1\}$, and where $\nu$ is a fixed distribution with support $\{0,1\}$. Using the same method as before (computing $\Psi_\epsilon^\mu$, then applying the concavification procedure), we establish the following result.

\begin{lem}
\label{lem:infoinc}
	For the stochastic network $\mathcal N_2$, full revelation of information is the unique optimal policy when $\epsilon\in(\sqrt{\nicefrac12},\sqrt{\nicefrac32})$, yet it is not optimal for $\epsilon\in(0,\sqrt{\nicefrac12}) \cup (\sqrt{\nicefrac32},\infty)$.
\end{lem}

This example illustrates that there may be no monotonicity of the information revealed in $\epsilon$ at all.

\section{Conclusion}

In this article, we studied the effect of bounded rationality on social welfare in the classic non-atomic routing game. Even though the price of anarchy---the usual metric to measure the inefficiency of equilibria resulting from aggregate selfish behavior---turns out inadequate when considering boundedly rational equilibria, we were able to define a new metric---the average excess time---which isolates the sole effect of bounded rationality. We have shown that the average excess time remains linear in $\epsilon$ when it is small, where the linearity constant (termed inefficiency factor) only depends on the topology of the network. This result is quite unexpected as the price of anarchy is famously independent of the topology. Our result establishes thus that topology plays an important role as soon as the time-minimizing assumption is slightly relaxed. This is perhaps best reflected at a fundamental level in the proof techniques. A  worst-case two-road network is leveraged to derive a price-of-anarchy inequality that stands for all topologies, whereas Theorem \ref{thm:continuity} relies on the analysis of the situation at hand for each possible near-UE support.

In this paper, we also investigated the effect of bounded rationality on a central planner's ability to mitigate traffic congestion using information design. The previous result proves most useful when $\epsilon$ is small, as it may be acceptable to use UE as an approximation for $\epsilon$-BRUE in designing an optimal signal. This is especially useful as the objective is harder to compute under the assumption drivers follow an $\epsilon$-BRUE rather than a UE. When $\epsilon$ is larger, however, the planner may wish to solve the program exactly. To study such cases, we turned towards specific instances to exhibit patterns in the optimal policy. We found that even though it is optimal to not reveal any information in the limit of large $\epsilon$, which was similarly observed by \citet{massicot2023almost} in the case of persuading a single $\epsilon$-maximizing agent for a quadratic game, there are instances where the information shared only increases, and others where no monotonicity exists.

In the future, we would like to extend this study to information design with richer communication channels such as private signals. The heterogeneous posterior belief environment makes the problem more challenging to tackle than public signals and the direct revelation principle---the usual information-design tool used to phrase the problem---fails to hold when agents are not exact utility-maximizers, however private signals offer more opportunity for persuasion and are thus worthy of consideration.

\bibliography{refs}             

\clearpage
                               
\appendix
\section{Proofs}    

\begin{pf}\emph{(of Proposition \ref{lem:varepsilon})}
	The first assertion holds trivially by definition. Regarding the second one, let $(f^n)$ be a sequence of $\mathcal F_d$ converging to some $f\in\mathcal F_d$. For all $n$ large enough, $\supp f^n \supset \supp f$ and thus,
	\begin{align*}
		\liminf_{n\to\infty} ~\varepsilon(f^n)
		&\geq \liminf_{n\to\infty} ~\max_{\substack{t\in\mathcal T\\P\in\mathcal P_t\cap\supp f\\Q\in\mathcal P_t}} ~c_P(f^n) - c_Q(f^n) \\
		&= \varepsilon(f).
	\end{align*}
	Regarding the function $\Psi_\cdot$, it is nondecreasing by definition and constant on $[\bar\epsilon,\infty)$ where
	\[
		\bar\epsilon = \max_{f\in\mathcal F_d} ~\|c(f)\|_\infty,
	\]
	since $\varepsilon(f) \leq \bar\epsilon$ for all $f\in\mathcal F_d$. The semicontinuity of $\Psi_\cdot$ is established as follows. The epigraph of $\varepsilon$ is closed by lower semincontinuity, moreover $\mathcal F_d$ is compact, hence the closed graph theorem asserts the constraint set correspondence $\epsilon\geq0\mapstto\{f\in\mathcal F_d, ~\varepsilon(f)\leq\epsilon\}$ is upper semicontinuous. This correspondence has nonempty compact values and $\Psi(\cdot)$ is continuous, thus the maximum theorem implies that $\Psi_\cdot$ is upper semicontinuous.
\end{pf}

\begin{pf}\emph{(of Proposition \ref{lem:variational})}
	Assume $f$ is an $\epsilon$-BRUE and let $g\in\mathcal F_d$. For all $t\in\mathcal T$ and $P\in\mathcal P_t$, let
	\[
		m_t = \min_{P\in\mathcal P_t} ~c_P(f), ~\Delta_P = c_P(f) - m_t\geq0.
	\]
	When $\Delta_P>\epsilon$, $f_P=0$ hence $f_P-g_P\leq0$ and so
	\begin{align*}
		\sum_{P\in\mathcal P} (f_P-g_P) c_P(f)
		&=\sum_{P\in\mathcal P} (f_P-g_P) \Delta_P\\
		&\leq 
		\sum_{t\in\mathcal T} 
		\sum_{\substack{P\in\mathcal P_t\\\text{ s.t. } \Delta_P\leq\epsilon}} (f_P-g_P) \Delta_P \\
		&\leq
		\sum_{t\in\mathcal T} 
		\sum_{\substack{P\in\mathcal P_t\\\text{ s.t. } \Delta_P\leq\epsilon}} \epsilon(f_P-g_P)^+ \\
		&= \epsilon \sum_{t\in\mathcal T} 
		\sum_{P\in\mathcal P_t} (f_P-g_P)^+ \\
		&= \frac12 \epsilon \|f-g\|_1.
	\end{align*}
	
	Conversely, let $f\in\mathcal F_d$ be a flow such that the statement holds. Consider then $t\in\mathcal T$ and $P,Q\in\mathcal P_t$ such that $f_P>0$. using the inequality with $g\in\mathcal F_d$ such that $g_P=0$, $g_Q=f_P+f_Q$ and $g_R=f_R$ otherwise,
	\[
		f_P (c_P(f) -c_Q(f)) \leq \epsilon f_P,
	\]
	hence
	\[
		c_P(f) \leq c_Q(f) + \epsilon.
	\]
\end{pf}

\begin{pf}\emph{(of Proposition \ref{lem:variational2})}
	Let $g$ be a UE and $f\in\mathcal F_d$. By convexity of $\Phi$,
	\begin{align*}
		\Phi(f) 
		&\leq \Phi(g) + \nabla\Phi(f)^\top (f-g)\\
		&= \Phi_0 + \sum_{P\in\mathcal P} c_P(f)(f_P-g_P) \\
		&\leq \Phi_0 + \frac12 \|f-g\|_1 \varepsilon(f)\\
		&\leq \Phi_0 + \|d\|_1 \varepsilon(f),
	\end{align*}
	where we used the result of Proposition \ref{lem:variational} for the second inequality.
\end{pf}

\begin{prop}
\label{thm:decoupled}
	Fix a network $(\mathcal G,c,d)$ with either constant latencies or with a single trip realized by disjoint paths. Let $f\in\mathcal F_d$ be a UE and $g\in\mathcal F_d$ a flow. For all $t\in\mathcal T$ and $P\in\mathcal P_t$ such that $g_P>0$,
	\[
		c_P(g) \leq \min_{Q\in\mathcal P_t} ~c_Q(f) + \varepsilon(g).
	\]
	In particular, $\psi \leq \varepsilon$, namely 
	\[
		\Psi(g) \leq \Psi_0 + \|d\|_1 \varepsilon(g).
	\]
\end{prop}

\begin{pf}\emph{(of Proposition \ref{thm:decoupled})}
	In the constant latencies case, for all $t\in\mathcal T$ and $P,Q\in\mathcal P_t$ such that $g_P>0$,
	\[
		c_P \leq c_Q + \varepsilon(g).
	\]
	Since $c$ is constant, the first conclusion applies. In the second case, the cost of path $P\in\mathcal P$ may be written simply in term of the flow that traverses it:
	\[
		c_P(f) = \sum_{e\in P} c_e(f_e) = \sum_{e\in P} c_e(f_P) = c_P(f_P).
	\]
	Let $g\in\mathcal F_d$ and let $f\in\mathcal F_d$ be a UE. Let then $Q\in\mathcal P$ be such that $f_Q>0$ and $g_Q \leq f_Q$. For all $P\in\mathcal P$ such that $g_P>0$,
	\begin{align*}
		c_P(g_P) 
		&\leq c_Q(g_Q) + \varepsilon(g)
		\leq c_Q(f_Q) + \varepsilon(g) \\
		&= \min_{R\in\mathcal P} ~c_R(f) + \varepsilon(g),
	\end{align*}
	thus the first statement applies. In both cases,
	\begin{align*}
		\Psi(g)
		&= \sum_{P\in\mathcal P} g_P c_P(g) \\
		&\leq \sum_{t\in\mathcal T} \sum_{P\in\mathcal P_t} g_P \left(\min_{Q\in\mathcal P_t} ~c_Q(f) + \varepsilon(g)\right) \\
		&= \sum_{t\in\mathcal T} d_t \left(\min_{Q\in\mathcal P_t} ~c_Q(f) + \varepsilon(g)\right)\\
		&= \Psi_0 + \|d\|_1 \varepsilon(g).
	\end{align*}
\end{pf}

\begin{pf}\emph{(of Theorem \ref{thm:continuity})}
	In the entirety of this proof, $\mathcal G=(\mathcal V,\mathcal E)$ is a fixed directed graph, $\mathcal T$ is the set of all possible non-vacuous trips and $\mathcal P$ is the set of all simple paths. We define the matrix $A$ by $A_{Pe} = \mathds 1_{e\in P}$ for all $e\in\mathcal E$ and $P\in\mathcal P$. We will rely on $A$ to define costs of paths and flows on edges. More precisely, if $f$ is a flow (on paths), it is understood as a vector of $\mathbb R^{\mathcal P}$ and the corresponding flow on edges is $A^\top f$. Likewise, if $\gamma$ is a cost vector (on edges), it is considered as an element of $\mathbb R^{\mathcal E}$ and the corresponding cost of each path is retrieved via the components of $A\gamma$. Nonetheless, when referring to \emph{components} of a vector, we may use the edge-wise and path-wise notations as found outside of this proof, namely $f_e$ refers to $(A^\top f)_e$ whereas $\gamma_P$ refers to $(A\gamma)_P$.
	
	Fix a network $(\mathcal G,c,d)$ for the moment and consider a sequence $(g^n)\in\mathcal F_d$ such that $\varepsilon(g^n)>0$ converges to $0$. By compactness of $\mathcal F_d$, we may assume without loss of generality that $g^n$ admits a limit $f\in\mathcal F_d$. The function $\varepsilon$ being lower semicontinuous, $\varepsilon(f)=0$, i.e., $f$ is a UE. Furthermore, since $2^{\mathcal P}$ is finite, we may assume that the support of $g^n$ is constant, we call it $\mathcal S'$.
	
	Call $\mathcal S$ the set of shortest paths that realize each trip in $\mathcal T$ under flow $f$. Define $\epsilon_0$ as the smallest time difference that separates a shortest path from a non-shortest path, i.e.,
	\[
		\epsilon_0 = \min_{t\in\mathcal T} 
		~\left(\min_{P\in\mathcal P_t\cap\mathcal S^\mathsf{c}} ~c_P(f)
		- \min_{P\in\mathcal P_t} ~c_P(f)\right) >0.
	\]
	When $n$ is large enough,
	\[
		\|A(c(g^n)-c(f))\|_\infty \leq \frac13\epsilon_0.
	\]
	For such $n$, if $\mathcal S' \not\subset \mathcal S$, then 
	\begin{align*}
		\varepsilon(g^n)
		&\geq \min_{t\in\mathcal T} 
		~\left(\min_{P\in\mathcal P_t\cap\mathcal S^\mathsf{c}} ~c_P(g)
		- \min_{P\in\mathcal P_t} ~c_P(g)\right) \\
		&\geq \frac13\epsilon_0>0.
	\end{align*}
	As a result, $\supp f \subset \mathcal S' \subset \mathcal S$. We may thus write
	\[
		\varepsilon(g^n)
		= \max_{\substack{t\in\mathcal T\\P\in\mathcal P_t\cap\mathcal S'\\Q\in\mathcal P_t}} ~(A\delta(g^n))_P - (A\delta(g^n))_Q,
	\]
	where $\delta(g) = c(g)-c(f) \in \mathbb R^{\mathcal E}$ for $g\in\mathcal F_d$. Moreover,
	\[
		\Psi(g^n) - \Psi_0
		= (g^n)^\top A \delta(g^n)
		\leq \|d\|_1 \vertiii{A}_\infty \|\delta(g^n)\|_\infty,
	\]
	where $\vertiii{A}_\infty$ refers to the operator norm of $A$ derived from the norm $\|.\|_\infty$.

	We may now temporarily forget the specific network $(\mathcal G,c,d)$ and $(g^n)$. The goal is to find a constant $\kappa>0$ such that
	\[
		\max_{\substack{t\in\mathcal T\\P\in\mathcal P_t\cap\mathcal S'\\Q\in\mathcal P_t}} ~(A\delta)_P - (A\delta)_Q
		\geq \kappa \|\delta\|_\infty
	\]
	for all $\delta\in\mathbb R^{\mathcal E}$ and $\mathcal S'$ that correspond to an actual instance. We will use the continuity of the left-hand side in $\delta$, the fact that it only vanishes at $\delta=0$ (using the essential uniqueness of UE), and the compactness of the set of $\delta$ corresponding to an instance, such that $\|\delta\|_{\infty} = 1$. 
	
	To define the latter set more precisely, fix $\mathcal S'\subset2^{\mathcal P}$. We let $\Sigma_{\mathcal S'}\subset\{-1,0,1\}^{\mathcal E}$ be the set of signs of $c(g)-c(f)\in\mathbb R^{\mathcal E}$ where $(\mathcal G,c,d)$ is a network, $f\in\mathcal F_d$ a UE, $\mathcal S$ the set of shortest paths for each trip and $g\in\mathcal F_d$ a non-UE flow with exact support $\mathcal S'$ such that $\supp f \subset \mathcal S' \subset \mathcal S$. We further define the compact set $\Delta_{\mathcal S'}$ as the set of $\delta\in\mathbb R^{\mathcal E}$ such that $\|\delta\|_\infty=1$ and such that there exists $\sigma\in\Sigma_{\mathcal S'}$ with the property that 1) $\sigma_e \delta_e \geq 0$ for all $e\in\mathcal E$ and 2) $\sigma_e = 0$ implies $\delta_e = 0$ for all $e\in\mathcal E$.
	
	Assume $\Delta_{\mathcal S'}$ is nonempty. The map $\eta_{\mathcal S'}$ defined by
	\[
		\eta_{\mathcal S'}(\delta) = \max_{\substack{t\in\mathcal T\\P\in\mathcal P_t\cap\mathcal S'\\Q\in\mathcal P_t}} ~(A\delta)_P - (A\delta)_Q \geq 0
	\]
	is continuous over the nonempty compact set $\Delta_{\mathcal S'}$, hence reaches a minimum $\kappa_{\mathcal S'}\geq0$ at some $\delta\in\Delta_{\mathcal S'}$. Let $\sigma\in\Sigma_{\mathcal S'}$ be such that $\sigma_e \delta_e \geq 0$ for all $e\in\mathcal E$, and $\sigma_e = 0$ implies $\delta_e = 0$ for all $e\in\mathcal E$. The sign vector $\sigma$ is the sign of $c(g)-c(f)\in\mathbb R^{\mathcal E}$ where $(\mathcal G,c,d)$ is a network, $f\in\mathcal F_d$ a UE, $\mathcal S$ the set of shortest paths for each trip and $g\in\mathcal F_d$ a non-UE flow with exact support $\mathcal S'$ such that $\supp f \subset \mathcal S' \subset \mathcal S$.
	
	Consider now an alternative network $(\mathcal G,c',d)$ where the costs $c'$ are any costs so that $c'(f) = c(f)$ and $c'(g) = c(f) + \lambda \delta$ where $\lambda>0$ remains to be fixed. To make sure such costs exist, we merely need to check that the definition of $c'$ at $f$ and $g$ satisfy for all $e\in\mathcal E$,
	\begin{enumerate}
		\item $c_e'(f_e)\geq0$,
		\item $c_e'(g_e)\geq0$,
		\item $(f_e-g_e) (c'_e(f_e)-c_e'(g_e)) \geq 0$,
		\item and $f_e=g_e$ implies that $c'_e(f_e)=c_e'(g_e)$.
	\end{enumerate}
	We check that each condition is satisfied:
	\begin{enumerate}
		\item Since $c_e$ is a nonnegative cost function, $c'_e(f_e) = c_e(f_e) \geq 0$. 
		\item This can be ensured using any $\lambda>0$ small enough as $c_e(f_e)=0$ and $\delta_e<0$ never occur simultaneously. Indeed, whenever $\delta_e<0$, $\sigma_e=-1$ and, as a result, $0\leq c_e(g_e) < c_e(f_e)$. 
		\item The condition is trivial when $c'_e(f_e)=c_e'(g_e)$. If $c'_e(f_e)-c_e'(g_e) = \lambda\delta_e>0$, then $\sigma_e=1$ so that $c_e(g_e)>c_e(f_e)$ and thus $g_e>f_e$. Similarly, if $c'_e(f_e)-c_e'(g_e) = \lambda\delta_e<0$, then $g_e<f_e$. Therefore, no matter the sign of $c'_e(f_e)-c_e'(g_e)$, (3) holds.
		\item If $f_e=g_e$, then $\sigma_e=0$, $\delta_e=0$ and, as a result, $c'_e(f_e)=c_e'(g_e)$.
	\end{enumerate}
	
	By definition of $c'$, $f$ is a UE of this second network. Since $c'(g) \neq c'(f)$, as $\lambda>0$ and $\delta\neq0$, $g$ is not a UE. As a result,
	\[
		0 
		< \varepsilon'(g)
		= \lambda \eta_{\mathcal S'}(\delta),
	\]
	thus $\kappa_{\mathcal S'} >0$. Let now $\kappa>0$ be a lower bound over all $\kappa_{\mathcal S'}$ such that $\Delta_{\mathcal S'}$ is not empty, and define $L_{\mathcal G} = \kappa^{-1}\vertiii{A}_\infty$. 
	
	Coming back to the first network $(\mathcal G,c,d)$ with sequence of non-UE $(g^n)$ converging to UE $f\in\mathcal F_d$, of constant support $\supp f\subset\mathcal S'\subset\mathcal S$, we have
	\[
		\varepsilon(g^n)
		= \eta_{\mathcal S'}(\delta(g^n))
		\geq \kappa \|\delta(g^n)\|_\infty,
	\]
	and thus,
	\[
		\Psi(g^n) - \Psi_0
		\leq \|d\|_1 \vertiii{A}_\infty \|\delta(g^n)\|_\infty
		\leq L_{\mathcal G} \|d\|_1 \varepsilon(g^n).
	\]

\bigskip

For the second claim, fix a network $(\mathcal G,c,d)$ and assume for the sake of contradiction that there exists a sequence of flows $(g^n) \in \mathcal F_d$ such that
\[
	\psi(g^n) > n \varepsilon(g^n).
\]
Since $\psi$ is continuous over the compact set $\mathcal F_d$, $\psi(g^n)$ is upper-bounded and thus $\varepsilon(g^n)$ vanishes. This is a contradiction since we have established that,
\[
	\psi(g) \leq L_{\mathcal G} \varepsilon(g) + o(\varepsilon(g)).
\]
\end{pf}

\begin{pf}\emph{(of Proposition \ref{prop:LG})}
Consider the network with $N$ chained Wheatstone networks as pictured in Figure \ref{fig:wheatstones}, where each Wheatstone network is a copy of the network of Figure \ref{fig:wheatstone}. The demand is set to $d'\in(0,1)$ on each subnetwork and $1-d'$ to cross all of them, hence the entire demand is
\[
	\|d\|_1 = Nd' + 1-d'.
\]
Let $f$ be the flow where the demand on each Wheatstone subnetwork is split into halves on up and down paths, and the demand to cross the entire network is split into one half using exclusively down paths and one using exclusively up paths. The cost of the paths of each subnetwork is $1$, hence $f$ is a UE. Consider the flow $g$ where a fraction $\nicefrac12-\eta$ of subnetwork flows routes on up and down paths and $2\eta$ use the middle path, and where the demand to cross the entire network still routes according to $f$. Inside each subnetwork, the cost of all up and down paths is $1+\eta d'$ whereas that of middle paths is $1+2\eta d'$. As a result, $\varepsilon(g) = \eta d'$ and
\[
	\Psi(g) - \Psi(f)
	= N \eta d' + 2 N (\eta d')^2.
\]
For this network then,
\[
	\mathcal L_{\mathcal G}
	\geq \frac{N}{Nd' + 1-d'}.
\]
This holding for all $d'\in(0,1)$, $\mathcal L_{\mathcal G} \geq N$.
\end{pf}

\begin{pf}\emph{(of Corollary \ref{cor:continuity})}
	For each $\mu\in\supp\tau$, Theorem \ref{thm:continuity} applied to the network $(\mathcal G,c^\mu,d)$ entails that
	\[
		\psi^\mu(g^\mu) \leq \mathcal L_{\mathcal G} \varepsilon^\mu(g^\mu) + o(\varepsilon^\mu(g^\mu)).
	\]
	Multiplying each of these finitely many inequalities by $\tau_\mu>0$ and summing them yields the desired statement.
\end{pf}

\begin{pf}\emph{(of Proposition \ref{lem:epsilonlarge})}
	For such $\epsilon$, the constraint $\varepsilon^\mu(g)\leq\epsilon$ is vacuous and thus the worst expected social cost is
\[
	\Psi^\mu_\epsilon = \max_{g\in\mathcal F_d} ~\Psi^\mu(g).
\]
For such $\epsilon$ then, $\Psi^\cdot_\epsilon$ is convex and it is therefore optimal to not reveal any information. 
\end{pf}

\begin{pf}\emph{(of Lemma \ref{lem:increase})}
	For $\epsilon\in[0,\nicefrac4{45}]$, the convexification of $\mu\in[0,1] \to \Psi^\mu_\epsilon$ is $\Psi_\epsilon$ itself on $[0,p(\epsilon)]$ and the tangent to $\Psi_\epsilon$ at $p(\epsilon)$ on $[p(\epsilon),1]$. This fact is directly derived from solving in $p$ for the equation
	\[
		\Psi_\epsilon'(p) (1-p) + \Psi_\epsilon(p) = \Psi_\epsilon(1).
	\]
	For $\epsilon\in[\nicefrac4{45},\nicefrac1{10}]$, the convexification of $\mu\in[0,1] \to \Psi^\mu_\epsilon$ is instead the line that passes through $(0,\Psi_\epsilon(0))$ and $(1,\Psi_\epsilon(1))$.
\end{pf}

\begin{pf}\emph{(of Lemma \ref{lem:infoinc})}
	The first consists in establishing $\Psi_\epsilon^\mu$. Plainly using the definition \eqref{eq:psiepsilon} and separating cases based on the support of $g\in\mathcal F_d$, we arrive at the following expressions. If $\epsilon\leq\nicefrac12$,
	\[
		\Psi_\epsilon^\mu = 
		\frac{\nicefrac32+\mu + \epsilon^2}2
		+ \frac{|\nicefrac12-\mu| \epsilon}2.
	\]
	If $\nicefrac12\leq\epsilon\leq1$,
	\[
		\Psi_\epsilon^\mu 
		=\begin{cases}
			1+\mu&\text{if }\mu\leq\epsilon-\nicefrac12\\
			\frac{\nicefrac32+\mu + \epsilon^2}2
			+ \frac{|\nicefrac12-\mu| \epsilon}2
			&\text{if }\epsilon-\nicefrac12\leq\mu\leq\nicefrac32-\epsilon\\
			\nicefrac32&\text{if }\nicefrac32-\epsilon\leq\mu.
		\end{cases}
	\]
	If $1\leq\epsilon\leq\nicefrac32$,
	\[
		\Psi_\epsilon^\mu 
		=\begin{cases}
			\frac{\nicefrac32+\mu - (\frac12-\mu) \epsilon + \epsilon^2}2&\text{if }\mu\leq\nicefrac32-\epsilon\\
			\max(\nicefrac32,1+\mu)&\text{if }\nicefrac32-\epsilon\leq\mu\leq\epsilon-\nicefrac12\\
			\frac{\nicefrac32+\mu + (\nicefrac12-\mu) \epsilon + \epsilon^2}2&\text{if }\epsilon-\nicefrac12\leq\mu.
		\end{cases}
	\]
	If $\epsilon\geq\nicefrac32$,
	\[
		\Psi_\epsilon^\mu =\max(\nicefrac32,1+\mu).
	\]
	
	When $\epsilon\in(\sqrt{\nicefrac12},\sqrt{\nicefrac32})$, the convexification of $\Psi_\epsilon^\mu$ is defined by the line that joins the points $(0,\Psi_\epsilon^0)$ and $(1,\Psi_\epsilon^1)$, hence full revelation is uniquely optimal. When $\epsilon\in(0,\sqrt{\nicefrac12}) \cup (\sqrt{\nicefrac32},\infty)$, the line that joins the points $(0,\Psi_\epsilon^0)$ and $(1,\Psi_\epsilon^1)$ lies strictly above the graph of $\mu\in[0,1]\to\Psi_\epsilon^\mu$ at $\nicefrac12$, and therefore full revelation is not optimal.
\end{pf}

\end{document}